\documentclass[prl,twocolumn,superscriptaddress,floatfix,nopacs,nofootinbib]{revtex4}
\usepackage{graphicx,amsfonts,amssymb,amsmath,hyperref,enumerate}
\usepackage[utf8]{inputenc}
\usepackage{xcolor}
\usepackage{tikz}
\usetikzlibrary{arrows.meta,
                chains,
                positioning,
                shapes.geometric}
\usepackage{siunitx}
\tikzstyle{startstop} = [rectangle, rounded corners, minimum width=3cm, minimum height=1cm,text centered, draw=black, fill=red!30]
\tikzstyle{io} = [trapezium, trapezium left angle=70, trapezium right angle=110, minimum width=3cm, minimum height=1cm, text centered, draw=black, fill=blue!30]
\tikzstyle{process} = [rectangle, minimum width=3cm, minimum height=1cm, text centered, draw=black, fill=orange!30]
\tikzstyle{decision} = [diamond, minimum width=3cm, minimum height=1cm, text centered, draw=black, fill=green!30]
\tikzstyle{arrow} = [thick,->,>=stealth]

\newif\ifhyper
\hypertrue
\ifhyper
\hypersetup{
   citecolor = {green},
   colorlinks = {true}, 
   urlcolor = {blue} 
}

\newcommand{\beq}{\begin{equation}}
\newcommand{\eeq}{\end{equation}}
\newcommand{\beqa}{\begin{eqnarray}}
\newcommand{\eeqa}{\end{eqnarray}}
\newcommand{\ket} [1] {\vert #1 \rangle}

\newcommand{\braket}[2]{\langle #1 | #2 \rangle}

\def\ket#1{\vert#1\rangle}

\def\Longarrow{\protect\@lra}
\def\@lra{\relbar\joinrel\relbar\joinrel\relbar\joinrel%
          \relbar\joinrel\rightarrow}

\begin{document} 

\title{Variational Quantum and Quantum-Inspired Clustering}

\author{Pablo Bermejo}
\affiliation{Multiverse Computing, Paseo de Miram\'on 170, E-20014 San Sebasti\'an, Spain}
\affiliation{Donostia International Physics Center, Paseo Manuel de Lardizabal 4, E-20018 San Sebasti\'an, Spain}

\author{Rom\'an Or\'us \footnote{Corresponding author: roman.orus@dipc.org}}
\affiliation{Multiverse Computing, Paseo de Miram\'on 170, E-20014 San Sebasti\'an, Spain}
\affiliation{Donostia International Physics Center, Paseo Manuel de Lardizabal 4, E-20018 San Sebasti\'an, Spain}
\affiliation{Ikerbasque Foundation for Science, Maria Diaz de Haro 3, E-48013 Bilbao, Spain}

\begin{abstract}
Here we present a quantum algorithm for clustering data based on a variational quantum circuit. The algorithm allows to classify data into many clusters, and can easily be implemented in few-qubit Noisy Intermediate-Scale Quantum (NISQ) devices. The idea of the algorithm relies on reducing the clustering problem to an optimization, and then solving it via a Variational Quantum Eigensolver (VQE) combined with non-orthogonal qubit states. In practice, the method uses  maximally-orthogonal states of the target Hilbert space instead of the usual computational basis, allowing for a large number of clusters to be considered even with few qubits. We benchmark the algorithm with numerical simulations using real datasets, showing excellent performance even with one single qubit. Moreover, a tensor network simulation of the algorithm implements, by construction, a quantum-inspired clustering algorithm that can run on current classical hardware.  

\end{abstract}

\maketitle

\emph{Introduction.-} Quantum computing is living interesting times. Right now, we are at the historical moment in which the first prototypes of quantum computers are allowing to think about actual applications. These first prototypes, called Noisy Intermediate-Scale Quantum (NISQ) devices, do not have error correction, are subject to noise, and have a limited number of qubits. Therefore, the relevant question is not ``when are we going to have a fully developed quantum computer?", but rather "can we do something useful with the limited machines that we have now?"

History has taught us that the answer to the above question must be a qualified \emph{yes}. One of the very first applications of quantum computers is \emph{quantum machine learning} \cite{Biamonte2017-hq}. This is a very active field of research, with new algorithms coming up every day. Among the many tasks in machine learning where a quantum computer can help one finds quantum neural networks, quantum support vector machines, and much more. All in all, there is a large variety of algorithms that have been proposed for supervised learning, where the quantum computer is trained with a set of data in order to recognize patterns and anomalies. A number of algorithms have been proposed in this context, including some that are amenable to the limitations of current quantum hardware such as quantum classifiers with data reuploading \cite{PerezSalinas2020datareuploading}. A different story, though, is that of  unsupervised learning, where the quantum computer must learn by itself the different classes associated to the data. These are the so-called \emph{clustering} algorithms, where computers cluster data into different groups according to the properties ``seen" by the algorithm. Concerning this, current approaches to quantum clustering rely on algorithms that are not easily implemented on NISQ devices. An example is the quantum-KNN algorithm \cite{qKNN}, which would require of more powerful quantum computers than current prototypes in order to apply it in a real-life setting (lots of data, lots of features, and perhaps highly imbalanced). Another option more recently explored is that of quantum annealing \cite{qannclus}, where clustering is mapped to an optimization. Despite being a friendlier algorithm, this option still has strong limitations, including the overhead due to embedding in the quantum annealer.

In this paper we propose an alternative approach to quantum clustering. Our idea is to solve it as an optimization problem, but on a universal gate-based quantum computer. This can be achieved by using a variational quantum circuit, i.e., by implementing a Variational Quantum Eigensolver (VQE). Moreover, we allow the target states in the Hilbert space to be non-orthogonal \cite{PerezSalinas2020datareuploading}. In this way we can implement a large number of clusters even if having few qubits. In practice, even just one single qubit is already able to implement an accurate clustering into many classes with just a single rotation in the Bloch sphere. By construction, this algorithm is perfectly adapted to run on current prototypes of universal gate-based quantum computers, and is able to deal with real, complex datasets. In addition, a tensor network \cite{tn1, tn2} simulation of the algorithm implements, by construction, a quantum-inspired clustering algorithm that can run on current classical hardware. And by targeting directly the optimization of the cost function with TNs, one has, directly, native quantum-inspired clustering algorithms. { The work presented in this paper is, therefore, not incremental: this is the first proposal of a quantum clustering algorithm that can solve real-size problems on gate-based NISQ quantum processors. This approach significantly reduces the computational requirements required for quantum clustering, so that it brings NISQ devices closer to practical applications of everyday life.}

\emph{Algorithm.-} Let us start by assuming that we have $N$ datapoints, each being described by $m$ features. The goal is to classify these datapoints into $k$ clusters. Without loss of generality, datapoints are described by $m$-dimensional vectors $\vec{x}_i$, with $i = 1, 2, \cdots, N$. To implement a clustering of the data we could, for instance, use classical bit variables $q_i^a = 0,1$, with $i = 1, 2, \cdots, N$ and $a = 1, 2, \cdots, k$, so that $q_i^a = 0$ if datapoint $i$ is not in cluster $a$, and $q_i^a = 1$ if it is in the cluster. Let us also call $d(\vec{x}_i, \vec{x}_j)$ some distance measure between datapoints $\vec{x}_i$ and $\vec{x}_j$.  With this notation we build a classical cost function $H$ such that points very far away tend to fall into different clusters \cite{qannclus}: 
\begin{equation}
    H=\frac{1}{2}\sum_{i,j=1}^N d(\vec{x}_i,\vec{x}_j)\sum_{a=1}^k q^a_iq^a_j. 
    \label{object}
\end{equation}
Additionally, one must impose the constraint that every point falls into one and only one cluster, i.e., 
\beq
\sum_{a=1}^k q^a_i = 1 ~~ \forall i. 
\eeq
The bit configuration optimizing Eq.(\ref{object}) under the above constraint provides a solution to the clustering of the data. As explained in Ref. \cite{qannclus}, this can be rephrased naturally as a Quadratic Binary Optimization Problem (QUBO) of $k \times N$ bit variables, so that it can be solved by a quantum annealer. However, on a gate-based quantum computer, we can use a Variational Quantum Eigensolver (VQE) \cite{vqe} with fewer qubits as follows. Let us call $f^a_i \equiv |\braket{\psi_i}{\psi^a}|^2$ the fidelity between a variational quantum state $\ket{\psi_i}$ for datapoint $\vec{x}_i$ and a reference state $\ket{\psi^a}$ for cluster $a$.  In a VQE algorithm, we could just sample terms $h_{ij}^a$,  
\beq
h_{ij}^a = d(\vec{x}_i,\vec{x}_j) f_i^a f_j^a, 
\label{cost0}
\eeq
for all datapoints $i, j$ and clusters $a$, together with penalty terms $c_i$, 
\beq
c_{i} = \left(\sum_{a=1}^k f_i^a - 1\right)^2, 
\label{cons}
\eeq
which are taken into account via Lagrange multipliers for all datapoints $i$. This last term must only be taken into account if several configurations of the qubits forming the VQE circuit allow for multiple clusters $a$ simultaneously for the same datapoint, e.g., if we codified one qubit per cluster as in Eq.(\ref{object}). 

Our approach here, though, is \emph{not to relate the number of qubits to the number of clusters}. Instead, we work with some set of predefined states $\ket{\psi^a} \in \mathcal{H}$, \emph{not necessarily orthogonal}, and being $\mathcal{H}$ whichever Hilbert space being used for the VQE. This provides us with enormous flexibility when designing the algorithm. For instance, we could choose states $\ket{\psi^a}$ to be a set of maximally mutually-orthogonal states \cite{PerezSalinas2020datareuploading} in $\mathcal{H}$. In the particular case of one qubit only, we would then have $\mathcal{H} = {\mathbb C}^2$ and the set of maximally-orthogonal states would correspond to the $k$ vertices of a platonic solid inscribed within the Bloch sphere.  The corresponding VQE approach would then correspond to a simple quantum circuit of just one qubit involving the fine-tuning of a single one-qubit rotation, and no sampling of the constraints in Eq.(\ref{cons}) would be needed at all, since this would be satisfied by construction. And for more qubits, the corresponding generalization would involve interesting entangled states in $\mathcal{H}$. 

In addition to this, the terms to be sampled can be further refined to improve algorithmic performance. One can for instance introduce modified cost functions, such as 
\beqa
h_{ij}^a &=& d(\vec{x}_i,\vec{x}_j)^{-1} \left(1 - f_i^a f_j^a \right)  \\
h_{ij}^a &=& \left( d(\vec{x}_i,\vec{x}_j)^\alpha + \lambda d(\vec{x}_i,\vec{c}_i)\right) f_i^a f_j^a   \\ 
h_{ij}^a &=& \left( d(\vec{x}_i,\vec{x}_j)^\alpha + \lambda d(\vec{x}_i,\vec{c}_i)\right) \left(1-f_i^a\right) \left(1- f_j^a \right). 
\label{cost2}
\eeqa
In the above cost functions, the first one tends to aggregate together in the same cluster those datapoints that are separated by a short distance, which is the complementary view to the original cost function in Eq.(\ref{cost0}). The second one includes two regularization hyperparameters $\alpha$ and $\lambda$, where $\alpha$ allows for modified penalizations for the distances between points, and $\lambda$ accounts for the relative importance of the distance between datapoint $\vec{x}_i$ and the centroid formed by the elements belonging to the same cluster than point $i$, which we call $\vec{c}_i$. This centroid can be re-calculated self-consistently throughout the running of the algorithm. Additionally, one can consider cost functions with a different philosophy, such as the third one, where datapoints with a large separation distance tend to be either in different clusters, but not ruling our the chance of being in the same cluster. On top of all these possibilities, one could also consider combining them in a suitable way to build even more plausible cost functions. Eventually, the goodness of a cost function depends on the actual dataset, so for each particular case it is worth trying several of them.  

The rest of the algorithm follows the standards in unsupervised learning. After a preprocessing of the data (e.g., normalization), we define the suitable set of states $\ket{\psi_a}$ and set the characteristics of the variational quantum circuit, including the parameters to be optimized. We set them the classical optimizer for the VQE loop (e.g., Adam optimizer) and its main features (learning rate, batch size...). After initialization, and if needed, we compute the centroids $\vec{c}_i$ and distances $d(\vec{x}_i, \vec{x}_j), d(\vec{x}_i, \vec{c}_i)$. We then perform the VQE optimization loop for a fixed number of epochs, where new parameters of the variational quantum circuit are computed at each epoch. To accelerate performance in the VQE loop, one can include only in the sampling those terms that have a non-negligible contribution. The final step involves estimating, for a given datapoint, the cluster to which it belongs. This can be done implementing quantum state tomography (either classical or quantum), so that we can read out the final state $\ket{\psi_i}$ for a given datapoint $\vec{x}_i$, and determine to which cluster it belongs by looking for the maximum of fidelities $f_i^a$ for all clusters $a$. 

\emph{Benchmark.-} We validated our approach by implementing  simulations both by generating random data distributed according to gaussian blobs, as well as using data from the Iris dataset \cite{Iris}. In our simulations we implemented all the cost functions mentioned in the previous section. In practice, we have seen that the one that shows better convergence and stability for the studied data was the last one in Eq.(\ref{cost2}). The results shown in this section correspond to that case.

\begin{figure}
\centering
\includegraphics[width=\linewidth]{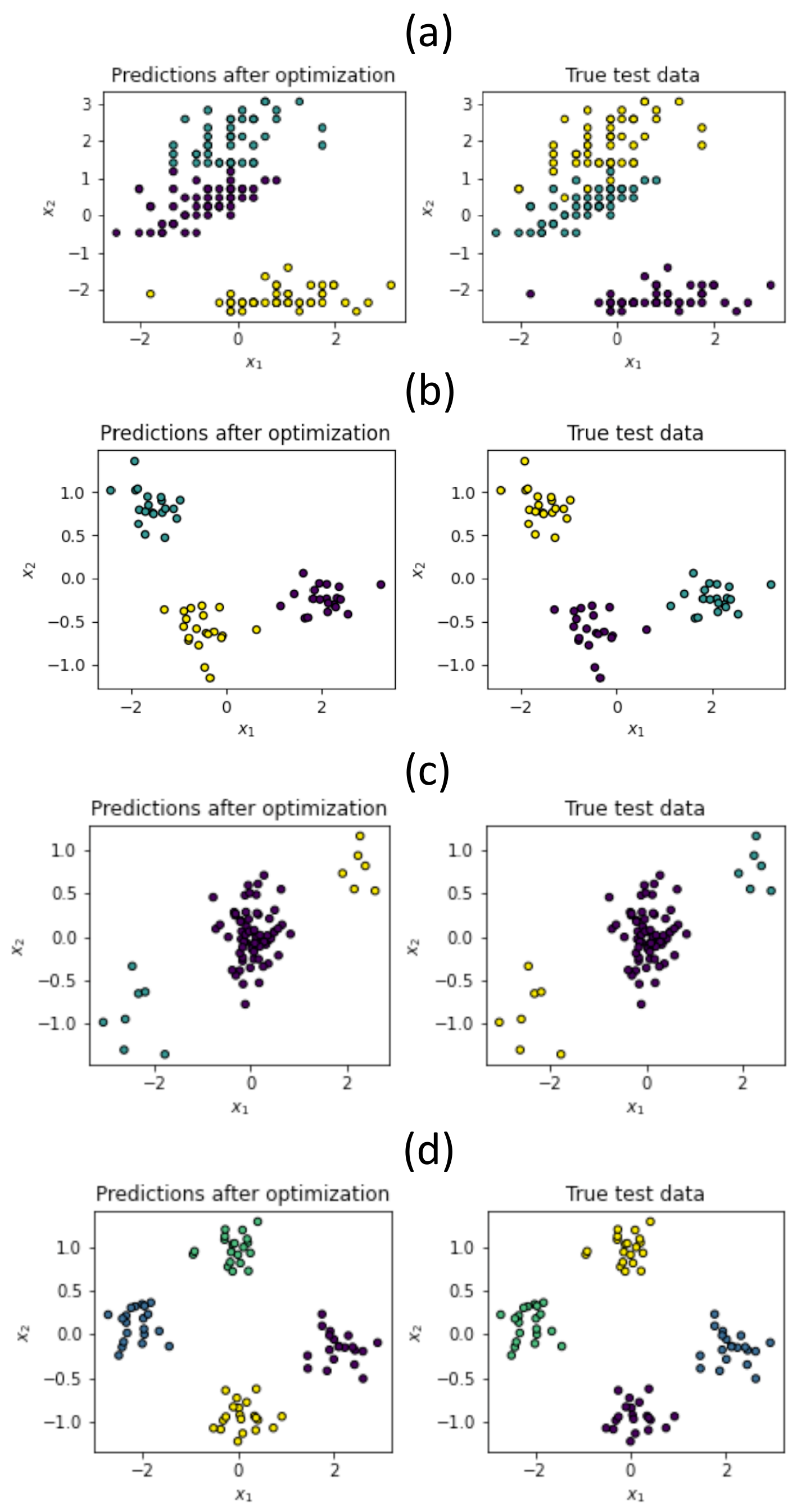}
\caption{[Color online] Clustering results for one qubit: (a) Iris dataset with three labels, (b) 3 gaussian blobs, (c) 3 gaussian blobs, (d) 4 gaussian blobs. Colors between true test and prediction do not necessarily match, since for the prediction the labelling is generated automatically by the algorithm.}
\label{fig1}
\end{figure}

\begin{figure}
\centering
\includegraphics[width=\linewidth]{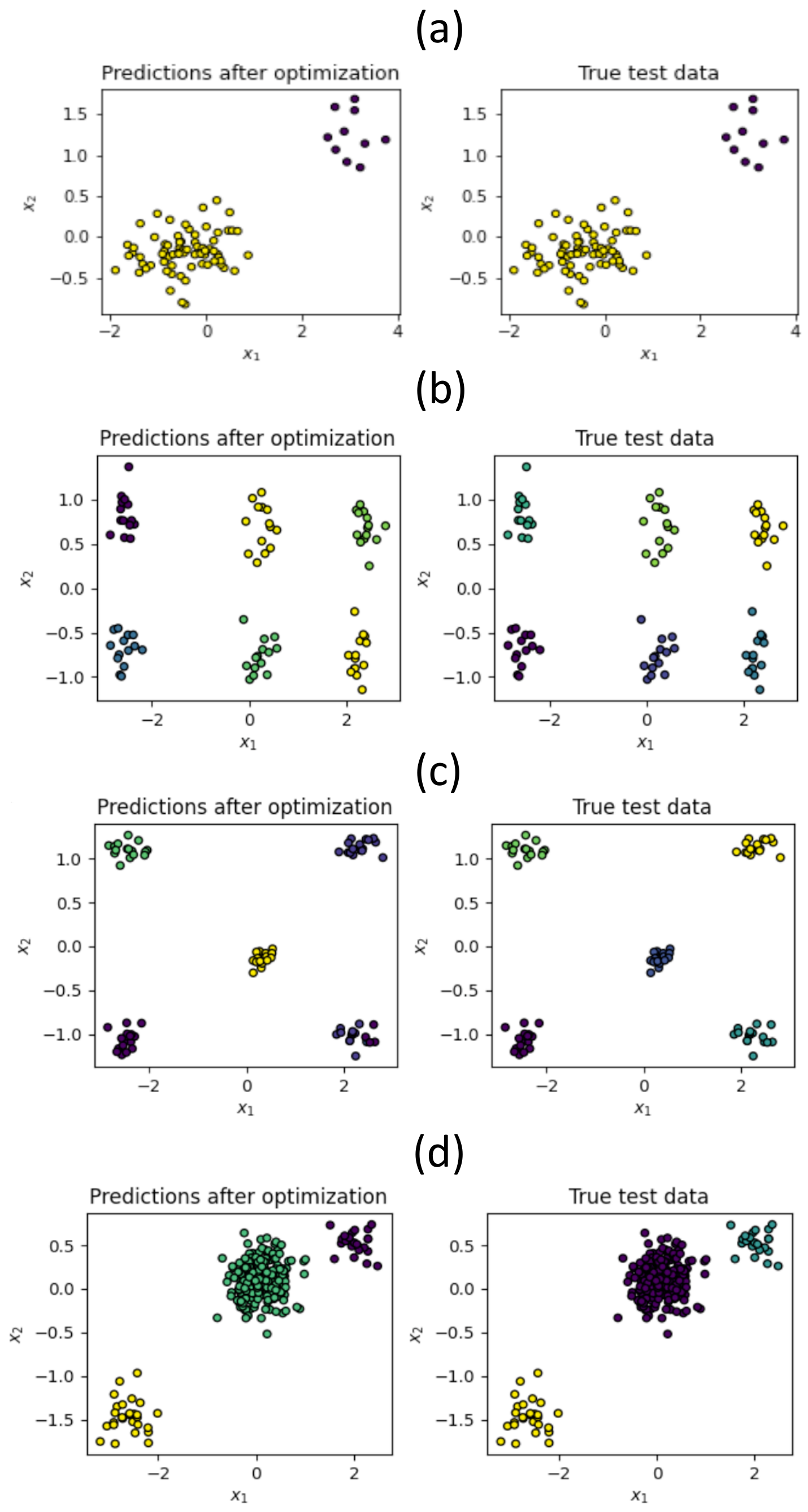}
\caption{[Color online] Clustering results for several qubits: (a) 2 qubits and 2 gaussian blobs, (b) 8 qubits and 6 gaussian blobs, (c) 10 qubits and 5 blobs, (d) 3 qubits and 3 gaussian blobs. Colors between true test and prediction do not necessarily match, since for the prediction the labelling is generated automatically by the algorithm.}
\label{fig2}
\end{figure}

In Fig.(\ref{fig1}) we show our results for one qubit. In Fig.(\ref{fig1}.(a)) we have the case of the Iris dataset. This dataset contains four features (length and width of sepals and petals) of 150 samples of three species of Iris (Iris setosa, Iris virginica and Iris versicolor). Here we take the classification in terms of sepal width vs petal with. The feature data is normalized and rescaled to two variables $x_1, x_2 \in [-1.9\pi/2,1.9\pi/2 ]$, which shows good algorithmic performance. In the left panel we show the results of the classification in the 3 classes using 15-20 epochs (iterations), achieving 96\% accuracy in the classification, as compared to the exact result which is shown in the right panel.   

\begin{figure}
\centering
\includegraphics[width=\linewidth]{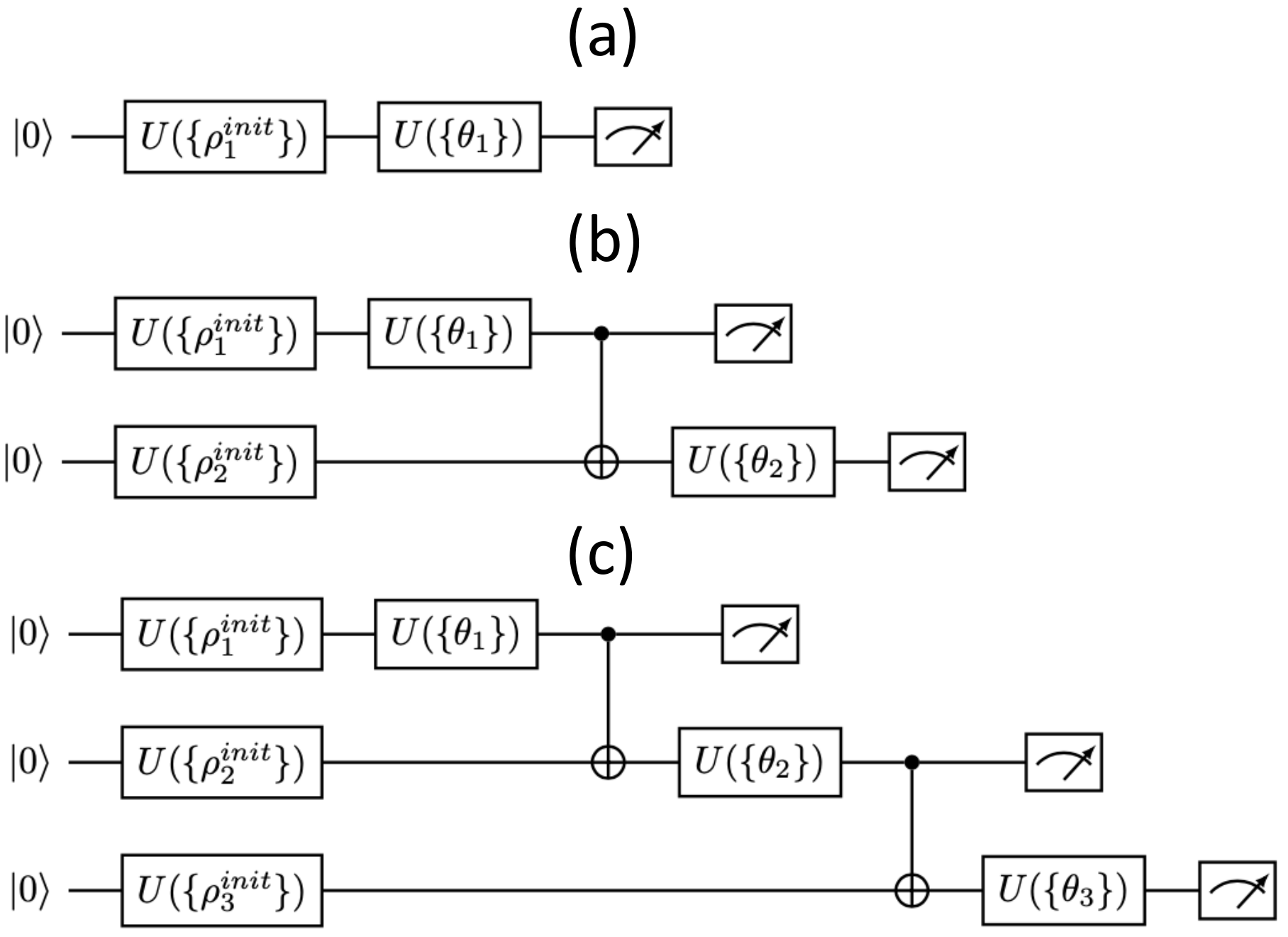}
\caption{Three variational quantum circuits for one epoch: (a) one qubit, (b) two qubits, and (3) three qubits. For qubit $i$, the set of variational angles is $\{ \theta_i \}$, and the set of angles for the initial rotations is $\{ \rho_i^{init} \}$.}
\label{fig3}
\end{figure}

In Fig.(\ref{fig1}.(b-d)) we show further results for one qubit, using gaussian blobs as data with 150-200 points for each case in a two-dimensional space with coordinates $x_1, x_2$. Again, the number of epochs was 15-20, and here in all cases the accuracy obtained was always 100\%. We see also that the higher the density of points, the better the algorithmic convergence and stability. 

Our results for several qubits are shown in Fig.(\ref{fig2}), for gaussian blobs and a similar scenario to that in the previous figure in terms of coordinates, epochs and accuracies, with increasing accuracy for increasing point density. Interestingly enough, in all the shown cases the Hilbert space allowed for more clusters to be considered. We see that in this case, an iterative strategy, increasing the number of clusters until we reach good accuracy, perfectly provides the number of clusters also without having to tell the algorithm a priori. Datasets with a larger number of clusters could also be considered.  

For the sake of explanation, in Fig.(\ref{fig3}) we include the idea of the variational quantum circuits that have been used for the different benchmarks, for 1, 2 and 3 qubits, in the case of one single epoch. For one qubit only the circuit involves only a single qubit rotation. In the case of two qubits, the most generic circuit involves two one-qubit rotations and one CNOT gate. For three qubits, we implement three one-qubit rotations and two sequential CNOTs. This strategy is also used for more qubits, namely for 8 and 10 qubits: one single-qubit rotation per qubit, and sequential entangling gates such as CNOTs and/or Toffolies. These circuits offer a good performance for the required data. What is more, they actually correspond to variational quantum states that are exactly Matrix Product States (MPS) \cite{tn1} of low bond-dimension. This has a double implication: first, it means that \emph{the same} clustering algorithm would work using tensor networks, in particular MPS, without having to use an actual quantum computer. And second, that for more complex and intricate datasets, we can always use more complex quantum circuits involving a large degree of entanglement between the qubits, so that they cannot be efficiently simulated with tensor networks (TN) \cite{tn1, tn2}. However, and as a matter fact, even a complex quantum circuit could be simulated using TNs at the expense of some controllable error, hence producing by construction a generic quantum-inspired clustering algorithm that can run on current classical hardware. In fact, with TNs one could target directly the optimization of the cost function without any quantum circuit behind, using e.g., variational, imaginary-time, or tangent space methods \cite{tn1, tn2}. 

{
\emph{Comparison to other approaches.-} Our approach has a number of advantages compared to alternative approaches to quantum clustering. Most prominently, our algorithm can be implemented reliably in a remarkably-small number of qubits. Moreover, it is a variational algorithm, and is therefore better suited to handle errors via techniques such as error mitigation. The fact that the optimization is heuristic allows also for extra degrees of freedom to play with, such as learning rates and, in fact, even the optimization algorithm (where one could play with gradient descent, stochastic gradient descent, adam optimizer, and more). As such, no other quantum clustering algorithm is so well adapted to run on NISQ devices. 

Let us review briefly the resources required by other quantum clustering approaches. Firstly, the original algorithm in Ref.\cite{horn} uses a Schroedinger approach, searching the potential $V(x)$ for which a a wavefunction $\psi(x)$ corresponds to estimator of the probability distribution of the data points. As such, this approach is an inverse optimization problem (i.e. finding the parent Hamiltonian of a given ground state), which is computationally hard and cannot be solved easily on a NISQ device. Secondly, the quantum-KNN algorithm \cite{qKNN} determines the distance between feature vectors encoded as quantum states, and for this uses Grover's quantum search algorithm \cite{grover} as a subroutine. The procedure therefore allows to obtain a square-root speedup with respect to the classical KNN method but, however, is remarkably hard to implement on few-qubit devices since Grover operators need larger quantum computers to be implemented in daily life scenarios. Hybrid approaches for quantum-KNN have also been proposed, see for instance Refs.\cite{hyb1, hyb2}. These approaches, while reducing in part the computational resources of the quantum-KNN alogorithm, still require the encoding of data into quantum states, which is difficult to implement and prone to errors. Additionally, the number of qubits required to run these algorithms is still large, since they still rely in part on Grover-like operators. Beyond quantum circuits, exotic approaches have also been proposed with other quantum computational models. For instance, in Ref.\cite{mbqc} the authors propose a quantum clustering algorithm for measurement-based quantum computers. Unfortunately, this model of quantum computer is not yet at the same level of hardware development  than the quantum circuit model. In addition, in Ref.\cite{qannclus}, the approach proposed based on quantum annealing needs a number of qubits for the annealer that is remarkably large and grows with the number of datapoints, therefore with unfavourable scaling.}

\emph{Conclusions.-} In this paper we have proposed a new quantum clustering algorithm that can run on NISQ devices. The algorithm is based on an optimization problem, that is subsequently solved via VQE using, on top, non-orthogonal states in the Hilbert space. The combination of all these approaches allows to cluster large datasets into a large number of clusters, accurately, and even with few qubits. The algorithm is benchmark by performing clusterings of the Iris dataset and gaussian blobs, using from one to ten qubits. Additionally, the algorithm can also be simulated and refined with tensor networks, producing a quantum-inspired clustering. 

We believe that the results in this paper are a significant step forward in the enormous challenge of finding useful applications of NISQ devices. As such, the proposed method can label real data, in an unsupervised way, even with few noisy qubits. The applications of such an algorithm are transversal in many fields of science, engineering and industry. Our algorithm shows that current prototypes of quantum computers can be applied in real-life settings, beyond toy academic models. 

\bigskip 

{\bf Acknowledgements.-} The authors acknowledge DIPC, Ikerbasque, Basque Government and Diputaci\'on de Gipuzkoa for constant support, as well as insightful discussions with the technical teams from Multiverse Computing, DIPC and MPC. 
\bigskip

{\bf Data availability.-} The datasets used and/or analysed during the current study are available from the corresponding author on reasonable request.

\bibliography{bibliography}
\bibliographystyle{apsrev4-1}

\end{document}